\documentstyle[prc,preprint,aps]{revtex}

\begin{document}
\preprint{}
\draft
 
\title{Comment on ``Validity of certain soft-photon amplitudes''}

\author{M.\ K.\ Liou$^a$, R.\ Timmermans$^b$, B.\ F.\ Gibson$^c$, 
        and Yi Li$^a$}
\address{\vspace{12pt}
$^a$Department of Physics and Institute for Nuclear Theory, 
         Brooklyn College of the City University of New York, 
         Brooklyn, New York 11210, USA\\
\vspace{12pt}
$^b$Kernfysisch Versneller Instituut,
         University of Groningen,  Zernikelaan 25,
         NL-9747 AA Groningen, The Netherlands\\
\vspace{12pt}
$^c$Theoretical Division, Los Alamos National Laboratory, 
         Los Alamos, New Mexico 87545, USA\\
}

\date{\today}

\maketitle

\begin{abstract}
The criteria suggested by Welsh and Fearing to judge the validity of 
certain soft-photon amplitudes are examined.  We comment on aspects of 
their analysis which lead to incorrect conclusions about published 
amplitudes and point out important criteria which were omitted from 
their analysis.
\end{abstract}

In a recent paper~\cite{Wel96} Welsh and Fearing discussed the 
validity of certain soft-photon amplitudes.  Their study focused on 
two issues, a phase space question and a symmetrization problem.
We comment on aspects of their arguments and procedures which lead 
to improper conclusions.

Evidence suggests that at least two different classes of soft-photon 
amplitudes are required to describe nuclear bremsstrahlung processes.  
The two-u-two-t special (TuTts) amplitudes, which represent a class 
of amplitudes evaluated using the Mandelstam variables 
($u_1, u_2, t_p, t_q$), were found to be optimal for processes 
involving strong u-channel exchange effects~\cite{Lio93,Lio95,Lio96}.  
The two-s-two-t special (TsTts) amplitudes, which represent a class 
of amplitudes evaluated at ($s_i, s_f, t_p, t_q$), were found to be
optimal for processes involving strong s-channel resonance 
effects~\cite{Lio93,Lio87,Din89}.  (The notation is defined in 
Ref.~\cite{Lio96}.)  These two classes of amplitudes can 
be derived for two bremsstrahlung cases: ({\it i}) 
$A + A \to A + A + \gamma$ and ({\it ii}) $A + B \to A + B + \gamma$.  
Alternative derivations were discussed in Ref.~\cite{Lio93,Lio96},  
but neither TsTts nor TuTts  amplitude has ever been defined 
for the third case ({\it iii}) $A + B \to C + D + \gamma$.  (Here 
$A, B, C,$ and $D$ represent particles having different masses and 
charges.)  The amplitudes given by Eqs.~(8), (9), and (11) of 
Ref.~\cite{Wel96} cannot be rigorously ``derived'' for the third 
case ({\it iii});  they were improperly ``generalized'' by the
authors. The danger of postulating rather than deriving results is 
further illustrated in Eq.~(31) of Ref.~\cite{Wel96}, where three 
invalid expressions for $\Delta_i$ are given.

The phase space problem related to the TsTts amplitude has already
been discussed in Ref.~\cite{Din89}.  Because the center-of-mass 
angle $\theta_{cm}$ cannot be defined for some kinematical points
involving ($s_f,t_p$) and ($s_f,t_q$), which correspond to those points 
that lie outside the measurable region shown in Fig.~2 of 
Ref.~\cite{Wel96}, the two-energy-four-angle version of the TsTts
amplitude cannot be used.  To circumvent this problem in 
Ref.~\cite{Lio95}, the special two-energy-two-angle (TETAS) 
amplitudes proposed in Ref.~\cite{Din89} were utilized for all TsTts 
calculations.

Because TuTts and TETAS amplitudes effectively describe different
bremsstrahlung processes, the theoretical constraints to be imposed 
upon them can differ.  For example, the TETAS amplitudes were shown 
to be the most successful in describing bremsstrahlung processes near 
a resonance~\cite{Lio87,Din89,Nef79,Lin91,Yan92}.  They satisfy an 
important criterion which was neglected in Ref.~\cite{Wel96}.  That 
is, in order to describe a bremsstrahlung process associated with a 
significant resonance, a valid amplitude must predict the correct 
(energy) position and width of the resonant peak, as observed in the 
bremsstrahlung spectrum.  Using Eqs.~(24) and (25) of Ref.~\cite{Lio87}, 
this criterion was investigated thoroughly~\cite{Lio87,Din89,Yan92}.
Processes like $\pi^+p\gamma$~\cite{Nef79,Lin91} and 
$p^{12}C\gamma$~\cite{Yan92} in the region of a resonance can only 
be well described by amplitudes which are evaluated at $s_i$ and 
$s_f$; the TETAS amplitudes were demonstrated to provide an excellent 
description of those processes.  The conventional Low amplitude 
fails to describe the $\pi^{\pm}p\gamma$ and $p^{12}C\gamma$ data 
in the vicinity of a resonance; in particular, it predicts 
incorrectly the position and width of the resonance peaks observed 
in the $p^{12}C\gamma$ spectrum~\cite{Lio87,Din89,Yan92}.

In Ref.~\cite{Wel96} it was concluded that ``the TuTts amplitude cannot 
be antisymmetrized while being written in terms of the measurable $pp$ 
elastic amplitude''.  Furthermore, it was stated that ``the failure 
in symmetrization arises at O(K/K)''.  We point out that this statement 
and conclusion are incorrect.  Firstly, the TuTts amplitude 
$M^{TuTts}_{\mu}$ ($\equiv M^{TuTts}_{1\mu}$) given by Eq.~(31) in 
Ref.~\cite{Lio96} {\it is} properly antisymmetrized, and it has been 
expressed in terms of the standard GGMW representation~\cite{Gol60} 
for the $pp$ elastic process.  This amplitude satisfies the Pauli
principle and other theoretical constraints.  Secondly, as shown in
Ref.~\cite{Lio96}, another TuTts amplitude $M^{TuTts}_{2\mu}$
[Eq.~(49)] can be obtained if the gauge invariance condition alone 
(without imposition of the Pauli principle) is used in the derivation.
This amplitude, which was used in Ref.~\cite{Lio95}, is not the
unique representative of the TuTts class.  The fact that this particular 
amplitude fails to satisfy the Pauli principle does not imply that the 
entire class of TuTts amplitudes must violate the Pauli principle.  
Finally, the amplitude $M^{TuTts}_{2\mu}$ violates the Pauli principle 
at order O(K), not at O(K/K) as stated in Ref.~\cite{Wel96}.  This can 
be seen by direct examination of the internal contribution to the
amplitude.  Alternatively, if one wishes to exhibit this violation
by comparing with the amplitude satisfying the Pauli principle,
then the comparison should be made with $M^{TuTts}_{\mu}$ and not the 
incorrect amplitude $M^{S\mu}_{TuTts}$  given by Eq.~(27) [or by 
subtracting Eq.~(44) from Eq.~(38)] of Ref.~\cite{Wel96}.  The same answer is 
obtained from either procedure.  That the violation is O(K) suggests
that the two amplitudes $M^{TuTts}_{\mu}$ and $M^{TuTts}_{2\mu}$ should 
predict very similar $pp\gamma$ cross sections except for kinematic 
conditions in which both proton scattering angles are very small and 
the photon angle $\psi_{\gamma}$ is near 180$^o$, which is borne out 
by numerical calculations.  As shown in Fig.~1, the coplanar 
$pp\gamma$ cross section at 157 MeV and for small symmetric proton 
scattering angles of 10$^o$ are almost the same for the two amplitudes 
except for $\psi_{\gamma}$ near 180$^o$.  It is also worth noting that 
the TRIUMF cross section data at 280 MeV, which disagree with some 
predictions calculated using the Low amplitude~\cite{Mic90,Lio95}, can 
actually be described satisfactorily by the $M^{TuTts}_{\mu}$ amplitude.

The incorrect conclusion drawn in Ref.~\cite{Wel96} follows directly
from the procedures employed in constructing a soft-photon amplitude.
These procedures, which differ from those utilized in Ref.~\cite{Lio93}
and \cite{Lio96}, involve three steps:  (A) Choose an unsymmetrized
bremsstrahlung amplitude $M_{\mu}$ which includes both the external
and internal contributions.  (B) Obtain an exchange amplitude 
($M_{\mu})_{p_f \leftrightarrow qf}$ by interchanging $p_f$ with 
$q_f$ (the momenta of the two outgoing protons) in $M_{\mu}$.  (C) 
Construct the symmetrized amplitude as 
$$
M^{S(A)} = M_{\mu} \pm (M_{\mu})_{p_f \leftrightarrow q_f} ,
$$
where the $S(A)$ corresponds to the $+(-)$ sign and describes two 
spin-0 (spin-$\frac{1}{2}$) particles.  Obviously, the initial step 
(A) is the crucial one.  The exchange amplitude 
($M_{\mu})_{p_f \leftrightarrow qf}$ can always be obtained from a 
given $M_{\mu}$, even when $M_{\mu}$ is not valid.  That is, a wrong 
amplitude $M_{\mu}$ leads to a wrong amplitude $M^{S(A)}$.  Step (A) 
specifies no detailed prescription for constructing $M_{\mu}$.  
Without a guiding prescription, an incorrect expression for $M_{\mu}$ 
can be chosen.  The treatment of the TuTts case in Ref.~\cite{Wel96} 
is a prime example.  Furthermore, Welsh and Fearing have emphasized 
that any soft-photon amplitude constructed should be written in terms 
of the corresponding elastic amplitude.  Certainly that is correct, 
but this condition is automatically satisfied if the procedures 
outlined in Ref.~\cite{Lio96} are followed.  (A properly 
antisymmetrized GGMW amplitude is the input for generating the 
$pp\gamma$ amplitude.)  However, a violation of this condition may 
arise when the procedures of Ref.~\cite{Wel96} are employed.

The procedures advocated in Ref.~\cite{Wel96} can be used to obtain 
the correct amplitude $M^{TuTts}_{\mu}$, provided that the procedures 
introduced in Ref.~\cite{Lio96} are utilized to generate a correct 
expression for $M_{\mu}$.  That is, $M_{\mu}$ should depend upon 
$V_{\mu}$ \{Eq.~(32) of Ref.~\cite{Lio96}\} and not $\bar{V}_{\mu}$ 
\{Eq.~(43) of Ref.~\cite{Lio96}\}.  In addition, the constraint 
given by Eq.~(7) of Ref.~\cite{Lio96} must be imposed in the 
derivation.


\vspace{6pt}
The work of M.K.L.\ was supported in part by the City University
of New York Professional Staff Congress-Board of Higher Education 
Research Award Program.  That of R.T.\ was included in the research
program of the Stichting voor Fundamenteel Onderzoek der Materie (FOM)
with financial support from the Nederlandse Organisatie voor
Wetenschappelijk Onderzoek (NWO).  That of B.F.G.\ was performed
under the auspices of the U.\ S.\ Department of Energy.

\vspace*{8pt}

\noindent Fig.\ 1.  Coplanar $pp\gamma$ cross section at 157 MeV 
for symmetric 10$^o$ proton angles.  The solid (symmetric) and 
dashed curves were calculated using the amplitudes $M^{TuTts}_{\mu}$ 
and $M^{TuTts}_{2\mu}$, respectively.

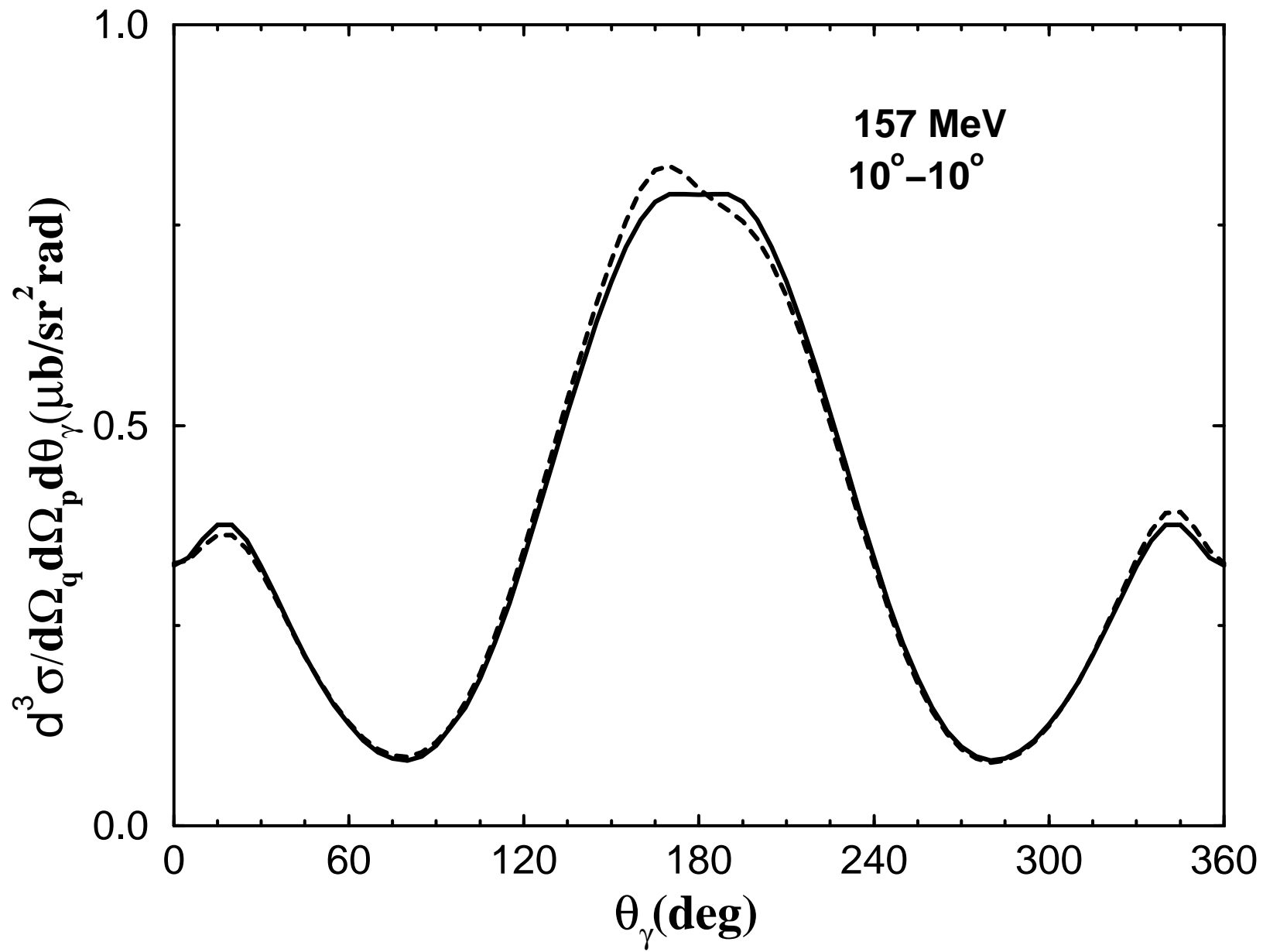
\end{document}